\documentclass[11pt]{article}
\usepackage[centering]{geometry}
\usepackage{amsmath,amssymb}
\usepackage{braket,subcaption,siunitx,booktabs,enumitem,mathtools,mathdots}
\usepackage{txfonts}
\usepackage{bm}
\usepackage{graphicx,xcolor}
\usepackage{comment}
\usepackage{cite}
\usepackage[unicode=true,bookmarks=true,bookmarksnumbered=false,bookmarksopen=false,breaklinks=false,pdfborder={0 0 0},pdfborderstyle={},backref=false,colorlinks=true]{hyperref}
\hypersetup{
 pdftitle={Extended supersymmetry in Dirac action with extra dimensions},
 pdfauthor={Yukihiro Fujimoto, Kouhei Hasegawa, Kenji Nishiwaki, Makoto Sakamoto, Kentaro Tatsumi, Inori Ueba},
 colorlinks=true,
 citecolor=blue
}
\newcommand{\one}{\boldsymbol{1}}

\newcommand{\Ncal}{\mathcal{N}}
\newcommand{\ebd}{\boldsymbol{e}}
\newcommand{\fbd}{\boldsymbol{f}}
\newcommand{\gbd}{\boldsymbol{g}}

\usepackage[T1]{fontenc}

\begin{document}

\title{
	\vspace{-30pt}
	{\Large \bf Extended supersymmetry with central charges\\ in higher dimensional Dirac action
		\\*[20pt]}
}

\author{
	Yukihiro~Fujimoto,$^a$\footnote{E-mail address: y-fujimoto@oita-ct.ac.jp} \quad
	Kouhei~Hasegawa,$^b$\footnote{E-mail address: kouhei@phys.sci.kobe-u.ac.jp} \quad
	Kenji~Nishiwaki,$^{c,d}$\footnote{E-mail address: knishiw@irb.hr} \\
	Makoto~Sakamoto,$^b$\footnote{E-mail address: dragon@kobe-u.ac.jp} \quad
	Kentaro~Tatsumi,$^b$\footnote{E-mail address: kentaro@stu.kobe-u.ac.jp} \quad
	Inori~Ueba,$^b$\footnote{E-mail address: i-ueba@stu.kobe-u.ac.jp}\\*[30pt]
	$^a${\it \normalsize National Institute of Technology, Oita College, Oita 870-0152, Japan}\\
	$^b${\it \normalsize Department of Physics, Kobe University, Kobe 657-8501, Japan}\\
	$^c${\it \normalsize Ru{\dj}er Bo{\v{s}}kovi{\'c} Institute, Division of Theoretical Physics, Zagreb 10000, Croatia}\\
	$^d${\it \normalsize School of Physics, Korea Institute for Advanced Study, Seoul 02455, Republic of Korea}\\*[55pt]
}

\date{
	\centerline{\small \bf Abstract}
	\begin{minipage}{0.9\textwidth}
		\medskip\medskip 
		\normalsize
A new realization of extended quantum-mechanical supersymmetry (QM SUSY) with central
extension is investigated.
We first show that two different sets of $d+2$ ($d+1$) supercharges for $d=$ even
(odd), each of which satisfies an $\mathcal{N}=d+2$ ($d+1$) extended QM SUSY
algebra without central extension, are hidden in the four-dimensional (4D)
mass spectrum of the $(4+d)$-dimensional Dirac action.
We then find that the whole set of the supercharges forms an $\mathcal{N}=2d+4$
($2d+2$) extended QM SUSY algebra with central charges for $d=$ even (odd).
The representation of the supersymmetry algebra is shown to be 
$1/2$-Bogomol'nyi--Prasad--Sommerfield states that correspond to a short representation
for the supersymmetry algebra with central extension.
We explicitly examine the 4D mass spectrum of the models with the hyperrectangle
and the torus extra dimensions, and discuss their supersymmetric structures.
		\begin{flushleft}
			\hspace{10ex}  extra dimensions
		\end{flushleft}
	\end{minipage}}

	\begin{titlepage}
		\maketitle
		\thispagestyle{empty}
	\end{titlepage}

\section{Introduction}
\label{sec:Introduction}

Quantum-mechanical supersymmetry (QM SUSY) is well known as the supersymmetry realized in quantum mechanics, which was introduced by Witten\cite{Witten1981} to investigate the supersymmetry breaking.
These days, the QM SUSY is applied to a wide range of research areas, e.g. exactly solvable systems in quantum mechanics \cite{Cooper, Quesne2005,Bagchi2009,Odake2009,Fernandez2009}, Berry phase \cite{Pedder:2007ff,Ohya:2014ska,Ohya:2015xya}, black holes and AdS/CFT \cite{claus1998,Gibbons1999,pacumio2000,Bellucci2003,Bakas2009}, Sachdev--Ye--Kitaev model \cite{Fu2017,Li2017,Yoon2017,Murugan2017,Jones2018}, extra dimensional models\cite{Lim2005,Lim2008a,Lim2008b,Nagasawa2011,Fujimoto2017b,Fujimoto2017a} and so on.
Recent trends in QM SUSY are reviewed in Ref.\cite{Fernandez2018}.

One of extensions of the QM SUSY is the $\mathcal{N}$-extended supersymmetry
\cite{Toppan2001,Kuznetsova2006,Faux2005,DeCrombrugghe1983,Howe1988,Akulov1999,Nagasawa2004,Nagasawa2005}
and another one is the central extension of the supersymmetry algebra \cite{Ivanov:1990cz,Niederle:2001cu,Faux:2003hn,Bellucci:2005gp}.
The $\Ncal$-extended supersymmetry has $\Ncal$ supercharges,
each of which corresponds to a square root of Hamiltonian, and they lead to
the degeneracy of the spectrum.
In addition, the central extension of supersymmetry is an extension that 
introduces central charges into the supersymmetry algebra \footnote{Spontaneous generations of the central charges in field-theoretic SUSY algebras and associated materials have been discussed (see e.g. \cite{Fayet:1975yi,Fayet:1978ig,Fayet:1984wm,Fayet:1985kt}).}.
Central charges are operators which commute with all the operators in the algebra,
and they can make the size of supermultiplets small, compared with 
the regular representation \cite{Witten:1978mh,WessBagger}. 
Such multiplets are called short multiplets or Bogomol'nyi--Prasad--Sommerfield (BPS) states\footnote{See also the original papers of BPS states \cite{Prasad:1975kr,Bogomolny:1975de}.}, 
and especially, $1/2$-BPS states are constructed from half of the supercharges.
Nevertheless in quantum mechanics, not so many models which realize arbitrary large 
$\Ncal$-extended QM SUSY with central charges are known.
Thus, it is worthwhile to investigate a new realization of the $\Ncal$-extended QM SUSY with central charges.

In Refs.\cite{Fujimoto2017b,Fujimoto2017a}, we have revealed that the $\mathcal{N}=2$
QM SUSY structure exists in the four-dimensional (4D) mass spectrum of 
the six-dimensional Dirac action.
Since a higher dimensional Dirac spinor can be decomposed into 4D Dirac
spinors with many \lq\lq flavors\rq\rq\ in the Kaluza--Klein (KK) 
decomposition,
we expect that some symmetries larger than the $\mathcal{N}=2$ QM SUSY
will be hidden in the 4D mass spectrum of higher dimensional Dirac actions.
Actually, it has been shown, in the previous paper \cite{Fujimoto:2018cnf},  that the 
$\mathcal{N}=2$ QM SUSY can be extended to the $\mathcal{N}= 2\lfloor d/2\rfloor+2$
QM SUSY (but without central extension) for the $(4+d)$-dimensional Dirac action,
where the symbol $\lfloor d/2\rfloor$ denotes the largest integer less than or 
equal to $d/2$.

In this paper, we discuss a new realization of $\mathcal{N}$-extended QM SUSY
with central extension.
Interestingly, we find another set of $2\lfloor d/2\rfloor+2$ supercharges
in the $(4+d)$-dimensional Dirac action, which forms the same 
$\mathcal{N}=2\lfloor d/2\rfloor+2$ QM SUSY algebra as that given in Ref.
\cite{Fujimoto:2018cnf}.
Then, we show that the whole set of the $4\lfloor d/2\rfloor+4$ supercharges
forms an $\mathcal{N}=4\lfloor d/2\rfloor+4$ extended QM SUSY algebra \textit{with}
central charges, and that the representation of the supersymmetry algebra forms
a short multiplet corresponding to $1/2$-BPS states.
We further verify that the supersymmetry clearly explains the structure of 
the 4D mass spectrum for the $(4+d)$-dimensional Dirac action 
in the hyperrectangle or the torus extra dimensions.
%

This paper is organized as follows: In Section \ref{sec:N=2-SUSY}, we summarize the KK decomposition of a $(4+d)$-dimensional Dirac field and show that the $\Ncal=2$
QM SUSY is hidden in the 4D mass spectrum. 
In Section \ref{sec:N-extended}, we give two sets of $2\lfloor d/2\rfloor+2$
supercharges, each of which satisfies the $2\lfloor d/2\rfloor+2$
QM SUSY algebra without central charges, and then show that the whole set of 
$4\lfloor d/2\rfloor+4$ supercharges forms the $\mathcal{N}=4\lfloor d/2\rfloor+4$
QM SUSY algebra with central charges.
In Section \ref{sec:rep-ex-SUSY}, we consider the representation of the algebra. 
Subsequently in Section \ref{sec:examples}, we examine concrete examples which realize 
the $\Ncal$-extended supersymmetry and confirm that the KK mode functions 
correspond to the representation given in Section \ref{sec:rep-ex-SUSY}. 
Section \ref{sec:summary} is devoted to summary and discussion.

%
\section{$\Ncal=2$ QM SUSY in higher dimensional Dirac action}
\label{sec:N=2-SUSY}
%
In this section, we give the setup of the $(4+d)$-dimensional Dirac action,
and review the structure of the $\mathcal{N}=2$ QM SUSY hidden in the 4D mass
spectrum of the system \cite{Fujimoto:2018cnf}.

Let us consider the $(4+d)$-dimensional Dirac action\footnote{
For earlier works on higher dimensional spinors, 
see e.g. \cite{Brauer1935,Wetterich1982,Kugo1982}.} 
with the 4D Minkowski space-time $M^{4}$ and a $d$-dimensional flat internal space $\Omega$:
%
\begin{align}
	S & =\int_{M^{4}}d^{4}x\int_{\Omega}d^{d}y\,\bar{\Psi}(x,y)
	(i\Gamma^{\mu}\partial_{\mu}+i\Gamma^{y_{k}}\partial_{y_{k}} - M \one_{2^{\lfloor d/2 \rfloor + 2}})\Psi(x,y)\,,
	\label{eq:action}
\end{align}
%
where the coordinates of $M^{4}$ and $\Omega$ are represented by $x^{\mu}$ ($\mu=0,1,2,3$) 
and $y_{k}$ ($k=1, 2, \cdots, d$), respectively.
The $\boldsymbol{1}_{2^{\lfloor d/2\rfloor+2}}$ denotes
the $2^{\lfloor d/2\rfloor+2}\times2^{\lfloor d/2\rfloor+2}$ unit matrix. The $\Gamma^{\mu}$ and $\Gamma^{y_{k}}$ indicate the 
$2^{\lfloor d/2\rfloor+2}\times2^{\lfloor d/2\rfloor+2}$ 
gamma matrices in $(4+d)$-dimensions and satisfy the Clifford algebra
%
\begin{align}
	\{\Gamma^A,\Gamma^B\}&=-2\eta^{AB}\one_{2^{\lfloor d/2\rfloor+2}}\quad 
	(A,B=0,1,2,3,y_{1},\cdots,y_{d})\,,
	\label{eq:Clifford_Definition}
\end{align}
%
where $\eta^{AB}$ is the $(4+d)$-dimensional metric defined by 
$\eta_{AB}=\eta^{AB}=\text{diag}(-1,+1,{\cdots},+1)$.
The parameter $M$ in the action \eqref{eq:action} is a bulk mass and $\Psi(x,y)$ is
a $(4+d)$-dimensional Dirac spinor with $2^{\lfloor d/2\rfloor+2}$ components.
The Dirac conjugate is defined as $\bar{\Psi}(x,y)=\Psi^{\dagger}(x,y)\Gamma^{0}$.

In this paper, we use the representation of the gamma matrices given by
the direct product of the internal spin space and the 4D one, i.e.\footnote{
We here adopt a slightly different representation of the gamma matrices from
that given in the previous paper \cite{Fujimoto:2018cnf}.
}
%
\begin{align}
	\Gamma^{\mu}=\one_{2^{\lfloor d/2\rfloor}}\otimes\gamma^\mu\,,&&
	\Gamma^{y_{k}}=\gamma^{y_{k}}\otimes\gamma^5\ \ (k=1,2,\cdots,d)\,,
	\label{eq:gamma_rep}
\end{align}
%
where $\gamma^{\mu}\ (\mu=0,\cdots,3)$ denote the $4\times4$ 4D gamma matrices and   
$\gamma^5\equiv i\gamma^0\gamma^1\gamma^2\gamma^3$ denotes the 4D chiral matrix.
The $\gamma^{y_{k}}$ ($k=1,2,\cdots,d$)
are the $2^{\lfloor d/2\rfloor}\times2^{\lfloor d/2\rfloor}$ internal space gamma matrices and satisfy
$\{\gamma^{y_k},\gamma^{y_l}\}=-2\delta^{kl}\one_{2^{\lfloor d/2\rfloor}}\,,\  (\gamma^{y_k})^\dagger=-\gamma^{y_k}\ (k,l=1,\cdots,d)\,$\footnote{
	For the case of $d=1$, we define $\gamma^{y_{1}}$ as $i$.}.

In terms of the 4D left-handed (right-handed) chiral spinors $\psi_{L,\alpha}^{(n)}(x)$
($\psi_{R,\alpha}^{(n)}(x)$), the KK decomposition of the
$(4+d)$-dimensional Dirac field $\Psi(x,y)$ will be given by
%
\begin{align}
	\Psi(x,y)
	& =\sum_{n}\sum_{\alpha}
	\Big\{\,\boldsymbol{f}_{\alpha}^{(n)}(y)\otimes\psi_{L,\alpha}^{(n)}(x)
	+\boldsymbol{g}_{\alpha}^{(n)}(y) 
	\otimes\psi_{R,\alpha}^{(n)}(x) \,\Big\}\,,
	\label{eq:expansion_Dirac}
\end{align}
%
where 
the index $n$ indicates the $n$-th level of the KK modes and $\alpha$ denotes 
the index that distinguishes the degeneracy of the $n$-th KK modes (if exists).
The mode functions $\boldsymbol{f}_{\alpha}^{(n)}(y)$ ($\boldsymbol{g}_{\alpha}^{(n)}(y)$)
have $2^{\lfloor d/2\rfloor}$ components and are assumed to form a complete set
with respect to the internal space associated with the 4D left-handed
(right-handed) chiral spinors $\psi_{L,\alpha}^{(n)}(x)$ 
($\psi_{R,\alpha}^{(n)}(x)$).

Substituting the expansion \eqref{eq:expansion_Dirac} into the action
\eqref{eq:action}, we require that the action can be written into the form
\begin{align}
	S=\int_{M^{4}}d^4x\,\bigg\{\,&\sum_{\alpha}\sum_{n}\bar{\psi}_{\alpha}^{(n)}(x)(i\gamma^{\mu}\partial_{\mu}-m_{n})\psi_{\alpha}^{(n)}(x)\nonumber\\
	&+\sum_{\alpha}\bar{\psi}_{L,\alpha}^{(0)}(x)i\gamma^{\mu}\partial_\mu\psi_{L,\alpha}^{(0)}(x)
	+\sum_{\alpha}\bar{\psi}_{R,\alpha}^{(0)}(x)i\gamma^{\mu}\partial_\mu\psi_{R,\alpha}^{(0)}(x)\,\bigg\}\,,
\end{align}
%
where $\psi_{\alpha}^{(n)}(x)=\psi_{L,\alpha}^{(n)}(x)+\psi_{R,\alpha}^{(n)}(x)$ indicate 4D Dirac spinors with mass $m_n$ and $\psi_{L/R,\alpha}^{(0)}(x)$ are massless 4D chiral spinors.
In order to obtain the above action, the mode functions $\boldsymbol{f}_{\alpha}^{(n)}(y)$
and $\boldsymbol{g}_{\alpha}^{(n)}(y)$ turn out to satisfy the following orthonormality
relations:
\begin{align}
	\int_{\Omega}d^{d}y\,\fbd_{\alpha}^{(n)\dagger}(y)\fbd_{\beta}^{({m})}(y)
	&=\int_{\Omega}d^{d}y\,\gbd_{\alpha}^{(n)\dagger}(y)\gbd_{\beta}^{({m})}(y)=\delta_{\alpha\beta}\delta^{nm}\,, 
	\notag\\
	\int_{\Omega}d^{d}y\,\fbd_{\alpha}^{(n)\dagger}(y)\big(A\gbd_{\beta}^{({m})}(y)\big)
	&=\int_{\Omega}d^{d}y\,\gbd_{\alpha}^{(n)\dagger}(y)\big(A^\dagger\fbd_{\beta}^{({m})}(y)\big)
	=m_{n}\delta_{\alpha\beta}\delta^{nm}\,,
	\label{eq:ortho}
\end{align}
where
$A=-i\gamma^{{y_k}}\partial_{{y_k}}+M \one_{2^{\lfloor d/2 \rfloor}}$ and 
$A^{\dagger}=i\gamma^{{y_k}}\partial_{{y_k}} + M \one_{2^{\lfloor d/2 \rfloor}}$\footnote{
It is noted that the definition of $A$ (and $A^\dagger$) is different from that in~\cite{Fujimoto:2018cnf} as
$$
- \Gamma^0 \left( i \, \Gamma^{y_k} \partial_{y_k} - M \one_{2^{\lfloor d/2 \rfloor + 2}} \right)
\equiv
\begin{pmatrix}
0 & A \\ A^\dagger & 0
\end{pmatrix}
\otimes \one_{2}.
$$
}.
Since the mode functions 
$\boldsymbol{f}_{\alpha}^{(n)}(y)$ and $\boldsymbol{g}_{\alpha}^{(n)}(y)$
are assumed to form the complete sets, the relations \eqref{eq:ortho} lead to
%
\begin{align}
	Q\begin{pmatrix}
	\boldsymbol{f}_{\alpha}^{(n)}(y) \\ 0
	\end{pmatrix}=m_n
	\begin{pmatrix}
	0 \\ \boldsymbol{g}_{\alpha}^{(n)}(y)
	\end{pmatrix}\,,&&
	H\begin{pmatrix}
	\boldsymbol{f}_{\alpha}^{(n)}(y) \\ 0
	\end{pmatrix}=m_n^2\begin{pmatrix}
	\boldsymbol{f}_{\alpha}^{(n)}(y) \\ 0
	\end{pmatrix}\,,&&
	(-1)^F\begin{pmatrix}
	\boldsymbol{f}_{\alpha}^{(n)}(y) \\ 0
	\end{pmatrix}=-\begin{pmatrix}
	\boldsymbol{f}_{\alpha}^{(n)}(y) \\ 0
	\end{pmatrix}\,,
\notag\\
	Q\begin{pmatrix}
	0 \\ \boldsymbol{g}_{\alpha}^{(n)}(y)
	\end{pmatrix}=m_n
	\begin{pmatrix}
	\boldsymbol{f}_{\alpha}^{(n)}(y) \\ 0
	\end{pmatrix}\,,&&
	H\begin{pmatrix}
	0 \\ \boldsymbol{g}_{\alpha}^{(n)}(y)
	\end{pmatrix}=m_n^2\begin{pmatrix}
	0 \\ \boldsymbol{g}_{\alpha}^{(n)}(y)
	\end{pmatrix}\,,&&
	(-1)^F\begin{pmatrix}
	0 \\ \boldsymbol{g}_{\alpha}^{(n)}(y)
	\end{pmatrix}=+\begin{pmatrix}
	0 \\ \boldsymbol{g}_{\alpha}^{(n)}(y)
	\end{pmatrix}\,,
	\label{eq:SUSY_relation}
\end{align}
%
where the supercharge $Q$, the Hamiltonian $H$ and the ``fermion'' number operator $(-1)^F$ are defined as
%
\begin{align}
	Q=\begin{pmatrix} 0  & A \\ A^{\dagger} & 0 \end{pmatrix}\,,&&
	H=Q^2=
	\begin{pmatrix}
	(-\partial_y^2+M^2) \one_{2^{\lfloor d/2 \rfloor}} & 0 \\ 0 & (-\partial_y^2+M^2) \one_{2^{\lfloor d/2 \rfloor}}
	\end{pmatrix}
	\,,&&
	(-1)^F=\begin{pmatrix}
	-\one_{2^{\lfloor d/2 \rfloor}} & 0 \\ 0 & \one_{2^{\lfloor d/2 \rfloor}}
	\end{pmatrix}\,.
\end{align}
%
The equations \eqref{eq:SUSY_relation} are nothing but the relations of 
$\Ncal=2$ supersymmetric quantum mechanics (SQM)\footnote{
If one defines $Q_{1}=Q$ and $Q_{2}=i(-1)^{F}Q$, then they form the $\Ncal=2$ SUSY algebra $\{Q_{j},Q_{k}\}=2H\delta_{jk}$ for $j,k=1,2$.}
\cite{Witten1981,Cooper}, and the mode functions 
$(\boldsymbol{f}_{\alpha}^{(n)}(y),0)^{\text{T}}$ 
and 
$(0,\boldsymbol{g}_{\alpha}^{(n)}(y))^{\text{T}}$ correspond to the ``bosonic'' 
and ``fermionic'' states that form an $\Ncal=2$ supermultiplet in the SQM.
Thus, we have found that the $\Ncal=2$ QM SUSY is hidden in the KK mode functions 
and the 4D mass spectrum. 

We should notice that the supercharge $Q$ has to be Hermitian to realize the 
$\mathcal{N}=2$ QM SUSY.
The Hermitian property of the supercharge $Q$ is assured if the KK mode functions
satisfy the condition for the surface integral
%
\begin{align}
	&\int_{\partial\Omega}d^{d-1}y\,
	\fbd_{\alpha}^{(n)\dagger}(y)\,
	in_{y_k}(y)\gamma^{y_k}
	\gbd_{\beta}^{({m})}(y) =0
	\label{eq:non-local_BC}
\end{align}
%
for all $m,n,\alpha,\beta$.
The $n_{y_k}(y)$ is a normal unit vector on the boundary $\partial\Omega$.
Since the above equation can be derived from the action principle $\delta S=0$,
the Hermiticity of the supercharge $Q$ is guaranteed as long as the Dirac field
obeys the action principle.
Thus, the $\Ncal=2$ QM SUSY is always realized in the 4D mass spectrum of the higher dimensional Dirac action and the doubly degenerate states $(\boldsymbol{f}_{\alpha}^{(n)}(y),0)^{\text{T}}$ and $(0,\boldsymbol{g}_{\alpha}^{(n)}(y))^{\text{T}}$ are mutually related 
by the supercharge $Q$, except for zero energy states.


\section{$\Ncal$-extended supersymmetry with central charges}
\label{sec:N-extended}
%
Although we have succeeded in explaining the degeneracy between the mode functions
$\boldsymbol{f}_{\alpha}^{(n)}(y)$ and $\boldsymbol{g}_{\alpha}^{(n)}(y)$
from an $\mathcal{N}=2$ supersymmetry point of view, we will see further degeneracy
labeled by $\alpha$ in the 4D mass spectrum.
We then  expect that some structures will be hidden furthermore in the 4D mass spectrum.
Actually, in the previous paper \cite{Fujimoto:2018cnf}, we have revealed that
an $\mathcal{N}=2\lfloor d/2\rfloor +2$ extended QM SUSY is hidden in the 4D
mass spectrum.

In this section, we first point out that there exist two sets of the supercharges, each of
which forms the $\mathcal{N}=2\lfloor d/2\rfloor +2$ extended QM SUSY algebra
without central extension.
We then show that the whole set of the supercharges satisfies the 
$\mathcal{N}=4\lfloor d/2\rfloor +4$ QM SUSY algebra \textit{with} central charges.
In the next subsection, we clarify the representation of the algebra.

\subsection{$\Ncal=2\lfloor d/2\rfloor+2$ supersymmetry for algebraic and geometric extensions}
\label{subsec:SUSY_alg_geo}
%
In this subsection, we explicitly construct two sets of supercharges, where one is
called algebraic and the other is geometric, respectively.
We then show that each set of them satisfies the $\mathcal{N}=2\lfloor d/2\rfloor +2$
supersymmetry algebra without central charges.
\\\\
{\bf  $\bullet$ Algebraic supercharges}

The $\mathcal{N}=2\lfloor d/2\rfloor +2$ supercharges based on the algebraic property 
of the gamma matrices are defined as follows:
%
\begin{align}
	Q_{k} & =\begin{pmatrix} 0 & -i\gamma^{d+1}\gamma^{y_k}A\\
	iA^{\dagger}\gamma^{d+1}\gamma^{y_k} & 0 \end{pmatrix}\,, & 
	Q_{d+1} & =Q\,, & 
	Q_{d+2} & =\begin{pmatrix} 0 & -i\gamma^{d+1}A\\
	iA^{\dagger}\gamma^{d+1} & 0 \end{pmatrix}\,, \\
	&(k=1,2,\dots,d)\,,   \notag
\end{align}
%
for $d=$ even, and
%
\begin{align}
	Q_{k} & =\begin{pmatrix} 0 & \gamma^{y_d}\gamma^{y_k}A\\
	-A^{\dagger}\gamma^{y_d}\gamma^{y_k} & 0
	\end{pmatrix}\,, & 
	Q_{d} & =Q\,,& 
	Q_{d+1} & =\begin{pmatrix} 0 & \gamma^{y_d}A\\
	-A^{\dagger}\gamma^{y_d} & 0 \end{pmatrix}\,, \\
	&(k=1,2,\dots,d-1)\,,   \notag
\end{align}
%
for $d=$ odd.
It should be noticed that
$\gamma^{d+1}\equiv i^{d/2}\gamma^{y_1}\cdots\gamma^{y_d}$ 
can be introduced only for $d=$ even and corresponds to the internal chiral matrix which satisfies
$\{\gamma^{d+1},\gamma^{y_k}\}=0\,,\ (\gamma^{d+1})^2= \boldsymbol{1}_{2^{\lfloor d/2\rfloor}}$ 
and $(\gamma^{d+1})^\dagger=\gamma^{d+1}$.
On the other hand, in the odd $d-$dimensions, one of $\gamma^{y_{k}}$ ($k=1,2,\cdots,d$) should be represented by the product of all the other gamma matrices. For the following sections, we use the representation of $\gamma^{y_d}=-i^{(d+1)/2}\gamma^{y_1}\cdots\gamma^{y_{d-1}}$ for $d=$ odd with $(\gamma^{y_d})^2= -\boldsymbol{1}_{2^{\lfloor d/2\rfloor}}$ and $(\gamma^{y_d})^\dagger=-\gamma^{y_d}$.

The above supercharges are found to satisfy the $\Ncal=2\lfloor d/2\rfloor+2$ supersymmetry algebra without central extension, i.e.
%
\begin{align}
	\{Q_i,Q_j\}=2H\delta_{ij}\,,&&[H,Q_i]=0 \ \ \ \ \ (i,j=1,2,\cdots,2\lfloor d/2\rfloor+2)\,.
\end{align}
%

\vspace{2mm}
\noindent
{\bf $\bullet$ Geometric supercharges}

Another set of the $\Ncal=2\lfloor d/2\rfloor+2$ supercharges can be constructed,
by use of the internal gamma matrices together with the reflection operators of the
internal space $\Omega$, as follows:
%
\begin{align}
	\tilde{Q}_{k} & =\begin{pmatrix} 0 & -i\gamma^{d+1}\gamma^{y_k}R_kA\\
	iA^{\dagger}\gamma^{d+1}\gamma^{y_k}R_k & 0 \end{pmatrix}\,, & 
	\tilde{Q}_{d+1}&=\begin{pmatrix} 0 & PA\\
	A^{\dagger}P & 0 \end{pmatrix}\,, &
	\tilde{Q}_{d+2} & =\begin{pmatrix} 0 & -i\gamma^{d+1}PA\\
	iA^{\dagger}\gamma^{d+1}P & 0 \end{pmatrix}\,,
	\label{eq:geo_even}\\
&(k=1,2,\dots,d)\,,\notag
\end{align}
%
for $d=$ even, and
%
\begin{align}
	\tilde{Q}_{k} & =\begin{pmatrix} 0 & \gamma^{y_d}\gamma^{y_k}R_dR_kA\\
	-A^{\dagger}\gamma^{y_d}\gamma^{y_k}R_dR_k & 0
	\end{pmatrix}\,, & 
	\tilde{Q}_{d}&=\begin{pmatrix} 0 & PA\\
	A^{\dagger}P & 0 \end{pmatrix}\,, &
	\tilde{Q}_{d+1} & =\begin{pmatrix} 0 & \gamma^{y_d}R_dPA\\
	-A^{\dagger}\gamma^{y_d}R_dP & 0 \end{pmatrix}\,,
		\label{eq:geo_odd}\\
&(k=1,2,\dots,d-1)\,,\notag
\end{align}
%
for $d=$ odd.
The $R_{k}$ ($k=1,2,\cdots,d$) represents  
the reflection operator for the $y_{k}$-direction\footnote{
The reflection operator $R_{k}$ ($k=1,2,\cdots,d$) is defined by
$(R_{k}f)(y_{1},\cdots,y_{k},\cdots,y_{d}) \equiv f(y_{1},\cdots,-y_{k},\cdots,y_{d})$
for any function $f(y_{1},\cdots,y_{d})$.
The $R_k$ and $\partial_l$ satisfy $R_k\partial_l =(1-2\delta_{kl})\partial_lR_k$.
}, 
and $P=\prod_{k=1}^{d}R_{k}$ denotes the point reflection
(or parity) operator of the internal space.
The above supercharges also realize the 
$\Ncal=2\lfloor d/2\rfloor+2$ supersymmetry algebra without central extension
%
\begin{align}
\{\tilde Q_i,\tilde Q_j\}=2H\delta_{ij}\,,&&[H,\tilde Q_i]=0 \ \ \ \ \ 
(i,j=1,2,\cdots,2\lfloor d/2\rfloor+2)\,.
\end{align}
%
It should be noted that the supercharges $\tilde{Q}_{i}$ are the same as 
those obtained in the previous paper \cite{Fujimoto:2018cnf}, 
except for $\tilde{Q}_{d+1}$ for $d=$ even and $\tilde{Q}_{d}$ for $d=$ odd.

\subsection{$\Ncal=4\lfloor d/2\rfloor+4$ supersymmetry with central charges}
\label{subsec:SUSY_with_CC}
%
In the previous subsection, we have seen that both of the  supercharges 
$Q_{i}$ and $\tilde{Q}_{i}$ ($i=1,2,\cdots,2\lfloor d/2\rfloor+2$)
satisfy the same $\mathcal{N}=2\lfloor d/2\rfloor+2$ QM SUSY algebra
without central extension.
Here, we show that the supercharges $Q_{i}$ together with 
$\tilde{Q}_{i}$ ($i=1,2,\cdots,2\lfloor d/2\rfloor+2$)
can extend the algebra to the $\mathcal{N}=4\lfloor d/2\rfloor+4$
QM SUSY algebra with central charges $(Z_i)$ such that
%
\begin{align}
	&\{Q_i,Q_j\}=
	\{\tilde{Q}_i,\tilde{Q}_j\}=2H\delta_{ij}\,,
	\notag\\
	&\{Q_i,\tilde{Q}_j\}=2Z_i\delta_{ij}\,,\\
	&[Z_i,Q_j]=[Z_i,\tilde{Q}_j]=[Z_i,Z_j]=[Z_i,H]=[H,Q_i]=[H,\tilde{Q}_i]=0\,,
	\notag\\
	&(i,j=1,2,\dots,2\lfloor d/2\rfloor+2)\,,\notag
\end{align}
%
where $Z_i$ ($i=1,2,\dots,2\lfloor d/2\rfloor+2$) are given by 
%
\begin{align}
	Z_k&= Q_k\tilde{Q}_k=
	\begin{pmatrix}
	A A^\dagger R_k & 0 \\ 0 & A^\dagger R_kA 
	\end{pmatrix}
	\ \ \ \ (k=1,2,\dots,d)\,,
	\notag\\
	Z_{d+1}&= Q_{d+1}\tilde{Q}_{d+1}=
	\begin{pmatrix}
	AA^\dagger P & 0 \\ 0 & A^\dagger PA 
	\end{pmatrix}\,,
	\label{eq:Z_01}
	\\
	Z_{d+2}&= Q_{d+2}\tilde{Q}_{d+2}=
	\begin{pmatrix}
	AA^\dagger P &  0 \\ 0 & A^\dagger PA 
	\end{pmatrix}\,, \notag
\end{align}
%
for $d=$ even, and
%
\begin{align}
	Z_k&= Q_k\tilde{Q}_k=
	\begin{pmatrix}
	AA^\dagger R_dR_k & 0 \\ 0 & A^\dagger R_dR_kA 
	\end{pmatrix}
	\ \ \ \ (k=1,2,\dots,d-1)\,,
	\notag\\
	Z_{d}&= Q_{d}\tilde{Q}_{d}=
	\begin{pmatrix}
	AA^\dagger P & 0 \\ 0 & A^\dagger PA
	\label{eq:Z_02}
	\end{pmatrix}\,,
	\\
	Z_{d+1}&= Q_{d+1}\tilde{Q}_{d+1}=
	\begin{pmatrix}
	AA^\dagger R_dP & 0 \\ 0 & A^\dagger R_dPA
	\end{pmatrix}\,, \notag
\end{align}
%
for $d=$ odd\footnote{
The supercharges especially satisfy 
$Q_{i}\tilde{Q}_{i}=\tilde{Q}_{i}Q_{i}\ \ (i=1,2,\cdots,2\lfloor d/2\rfloor+2)\,.$}.
Since $Z_i$ ($i=1,2,\cdots,2\lfloor d/2\rfloor+2$) commute with all the operators, 
we can regard $Z_i$ as central charges in this algebra. As we can see the form 
of the central charges, they may be interpreted as the reflection operators 
(accompanied with the Hamiltonian) compatible with the QM SUSY.

In the next section, we discuss the representation of the supersymmetry algebra.
For this purpose, it is convenient to adopt the basis of supercharges as
%
\begin{align}
	&Q_i^{\pm}=\frac{1}{2}(Q_i\pm\tilde{Q}_i)\ \ \ \ (i=1,2,\cdots,2\lfloor d/2\rfloor+2)
	\label{eq:redef_supercharge}\,.
\end{align}
%
The explicit forms of the supercharges $Q^{\pm}_{i}$ are given as 
%
\begin{align}
	Q_{k}^\pm & =\begin{pmatrix} 0 & -i\gamma^{d+1}\gamma^{y_k}\Pi_k^\pm A\\
	iA^{\dagger}\gamma^{d+1}\gamma^{y_k}\Pi_k^\pm & 0 \end{pmatrix}\,, & 
	Q_{d+1}^\pm & =\begin{pmatrix} 0 & \Pi_{d+1}^\pm A\\
	A^{\dagger}\Pi_{d+1}^\pm & 0 \end{pmatrix}\,, &
	Q_{d+2}^\pm & =\begin{pmatrix} 0 & -i\gamma^{d+1}\Pi_{d+1}^\pm A\\
	iA^{\dagger}\gamma^{d+1}\Pi_{d+1}^\pm & 0 \end{pmatrix}\,,
	  \notag\\
	&(k=1,2,\dots,d)\,, 
	\label{eq:Qpm_even}
\end{align}
%
for $d=$ even, and
%
\begin{align}
	Q_{k}^\pm & =\begin{pmatrix} 0 & \gamma^{y_d}\gamma^{y_k}\Pi_{dk}^\pm A\\
	-A^{\dagger}\gamma^{y_d}\gamma^{y_k}\Pi_{dk}^\pm & 0 \end{pmatrix}\,, & 
	Q_{d}^\pm & =\begin{pmatrix} 0 & \Pi_{d+1}^\pm A\\
	A^{\dagger}\Pi_{d+1}^\pm & 0 \end{pmatrix}\,, &
	Q_{d+1}^\pm & =\begin{pmatrix} 0 & \gamma^{y_d} \Pi_{d(d+1)}^\pm A\\
	-A^{\dagger}\gamma^{y_d}\Pi_{d(d+1)}^\pm & 0 \end{pmatrix}\,, \notag\\
	&(k=1,2,\dots,d-1)\,, 
	\label{eq:Qpm_odd}
\end{align}
%
for $d=$ odd, where $\Pi_{k}^\pm=(1\pm R_k)/2\,,\Pi_{d+1}^\pm=(1\pm P)/2\,,\Pi_{dk}^\pm=(1\pm R_dR_k)/2$
and $\Pi_{d(d+1)}^\pm=(1\pm R_dP)/2$ play the role of the projection operators.
Then, these supercharges are found to satisfy the algebra
%
\begin{align}
	&\{Q_i^{\pm},Q_j^{\pm}\}=(H\pm Z_i)\delta_{ij}\,,
	\label{eq:redef_SUSY_alg1}\\
	&\{Q_i^{\pm},Q_j^{\mp}\}=0\,,
	  \hspace{15mm} (i,j=1,2,\cdots,2\lfloor d/2\rfloor+2)\,.
	\label{eq:redef_SUSY_alg2}
\end{align}
%
In the next section, we use this basis of the supercharges.

It should be pointed out that all of the supercharges $Q_{i}$ and $\tilde{Q}_{i}$
(or $Q^{\pm}_{i}$)\ ($i=1,2,\cdots,2\lfloor d/2\rfloor+2$)
would not be necessarily well-defined in the system.
In order for them to be well defined, 
the supercharges $Q_{i}$ and $\tilde{Q}_{i}$ have to be compatible with
boundary conditions (if the internal space $\Omega$ has boundaries), that is, 
for any state $\Phi(y)$, $Q_{i}\Phi(y)$ and
$\tilde{Q}_{i}\Phi(y)$ should obey the same boundary condition as that on $\Phi(y)$,
otherwise the action of $Q_{i}$ and $\tilde{Q}_{i}$ on $\Phi(y)$ is ill-defined.
Furthermore, in order for $\tilde{Q}_{i}$ to be well-defined, the reflection
operators $R_{k}$\ ($k=1,2,\cdots,d$) should properly act on the internal space
$\Omega$.
In this paper, we restrict our considerations to the cases that all the supercharges
$Q_{i}$ and $\tilde{Q}_{i}$ (or $Q_{i}^{\pm}$) ($i=1,2,\cdots,2\lfloor d/2\rfloor+2$)
are well-defined with the $\mathcal{N}=4\lfloor d/2\rfloor+4$ QM SUSY algebra.

\section{Representation of $\Ncal$-extended supersymmetry with central charges}
\label{sec:rep-ex-SUSY}

In this section, we clarify the representation of the supersymmetry algebra
derived in the previous section for the non-zero energy states.

Since the Hamiltonian $H$ and the central charges $Z_{i}$
($i=1,2,\cdots,2\lfloor d/2\rfloor+2$) commute with each other, we can introduce
the simultaneous eigenstates of $H$ and $Z_{i}$.
Furthermore, since the central charges satisfy the relations
%
\begin{align}
(Z_i)^2=H^2 \qquad (i=1,2,\cdots,2\lfloor d/2\rfloor+2)\,,
\label{eq:rel_Z^2=H^2}
\end{align}
%
with
%
\begin{align}
&H^{d-1}Z_{d+1}=H^{d-1}Z_{d+2}= Z_{1}Z_{2}\cdots Z_{d}\,,
  \hspace{10mm} \text{for}\ d=\text{even}\,,
\label{eq:rel_Z_even} \\
& H^{d-2}Z_{d+1}= Z_{1}Z_{2}\cdots Z_{d-1}\,,
  \hspace{26mm} \text{for}\ d=\text{odd}\,,
\label{eq:rel_Z_odd}
\end{align}
%
the eigenvalues of $H$ and $Z_{i}$ for the non-zero energy states with
$m_{n}^{2} \ne 0$ can be parameterized as follows:
%
\begin{align}
H\,\Phi_{\alpha,\,\vec{z}\,}^{(n)}(y) 
 &= m_{n}^{2}\, \Phi_{\alpha,\,\vec{z}\,}^{(n)}(y) \,, \\
Z_{i}\,\Phi_{\alpha,\,\vec{z}\,}^{(n)}(y) 
 &= z_{i}\,m_{n}^{2}\, \Phi_{\alpha,\,\vec{z}\,}^{(n)}(y)\,,
   \quad (i=1,2,\cdots,2\lfloor d/2\rfloor+2)\,,
\end{align}
%
with\footnote{
It should be noted that the eigenvalues $z_{i}=\pm 1$ can be defined 
without ambiguity for the non-zero energy states with $m_{n}^{2} \ne 0$, 
and also that all of the eigenvalues $z_{i}$
($i=1,2,\cdots,2\lfloor d/2\rfloor+2$) are not independent but only
$z_{k}$ ($k=1,2,\cdots,d$) are independent. 
}
%
\begin{align}
&z_{i} = \pm 1 \qquad (i=1,2,\cdots,2\lfloor d/2\rfloor+2)\,, \\
&z_{d+1} = z_{d+2} = z_{1}z_{2}\cdots z_{d} \qquad  \text{for } d = \text{even}\,,
    \label{eq:dependent_z1}\\
&z_{d+1} = z_{1}z_{2}\cdots z_{d-1} \qquad\qquad \text{for } d = \text{odd}
    \label{eq:dependent_z2}\,,
\end{align}
%
where $\vec{z} = (z_{1},z_{2},\cdots,z_{d})$
and the index $\alpha$ labels the degeneracy for fixed $m_n$ with $\vec{z}$ \footnote{
The label $\alpha$ given in Eq. \eqref{eq:expansion_Dirac} corresponds to \{$\alpha,\vec{z}$\} defined in this section.}.

It may be worth while explaining physical meanings of the discrete eigenvalues $z_{i}=\pm1$.
It follows from the expressions \eqref{eq:Z_01} and \eqref{eq:Z_02} that
the central charges $Z_{i}$ essentially correspond to the reflection operators 
(accompanied with the Hamiltonian), so that $z_{i}$ may be interpreted as 
the labels for \lq\lq parity\rq\rq\ even or odd of the eigenfunctions.
It should be, however, emphasized that the reflection operators $R_{k}$
($k=1,2,\cdots,d$) and $P$ themselves do not commute with the supercharges
$Q_{i}^{\pm}$ and hence they are not compatible with the supersymmetry.
On the other hand, $Z_{i}$ commutes with all the supercharges, so that
$Z_{i}/H$ (for the non-zero energy states) can be regarded as a
\lq\lq reflection\rq\rq\ operator compatible with the supersymmetry.

In order to construct the representation, i.e. the supermultiplet of the
supersymmetry algebra \eqref{eq:redef_SUSY_alg1}, \eqref{eq:redef_SUSY_alg2},
we first note that $Q_{i}^{-z_{i}}$ ($i=1,2,\cdots,2\lfloor d/2\rfloor+2$)
acts trivially on $\Phi_{\alpha,\,\vec{z}\,}^{(n)}(y)$ with
$\vec{z}=(z_{1},z_{2},\cdots,z_{d})$, i.e.
%
\begin{align}
Q_i^{-z_i}\,\Phi^{(n)}_{\alpha,\,\vec{z}\, }(y)=0
  \quad (i=1,2,\cdots,2\lfloor d/2\rfloor+2)\,.
  \label{eq:QPhi=0}
\end{align}
%
This is because the relations \eqref{eq:redef_SUSY_alg1} implies
$
\big(Q_i^{-z_i}\big)^{2}
\Phi^{(n)}_{\alpha,\,\vec{z}\, }(y)=0$,
which leads to \eqref{eq:QPhi=0} due to the Hermitian property of the 
supercharges.
Thus, the supercharges that act on the states
$\Phi_{\alpha,\,\vec{z}\,}^{(n)}(y)$ non-trivially are given by the set of
$\{\,Q_{i}^{z_{i}}\ (i=1,2,\cdots,2\lfloor d/2\rfloor+2)\,\}$
and the number of the supercharges turns out to reduce effectively to half\footnote{
Note that the set of the non-trivial supercharges 
$\Big\{\,Q_{i}^{z_{i}}\ (i=1,2,\cdots,2\lfloor d/2\rfloor+2)\,\Big\}$
depends on the eigenvalues $\vec{z}=(z_{1},z_{2},\cdots,z_{d})$ 
of the state $\Phi^{(n)}_{\alpha,\,\vec{z}\,}(y)$.
}.

The supermultiplet associated with the state
$\Phi_{\alpha,\,\vec{z}\,}^{(n)}(y)$
can be constructed in the following way.
In terms of the non-trivial supercharges $Q_{i}^{z_{i}}$, it will be useful
to introduce the operators 
%
\begin{align}
	S_p^{z_{2p-1}z_{2p}}= -iQ_{2p-1}^{z_{2p-1}} Q_{2p}^{z_{2p}} 
	 \qquad (p=1,2,\cdots,\lfloor d/2\rfloor+1)\,.
\end{align}
%
They are explicitly given by
%
\begin{align}
&S_p^{z_{2p-1}z_{2p}}
  =	\begin{pmatrix}
	AA^\dagger \gamma_{(p)}\Pi_{2p-1}^{z_{2p-1}}\Pi_{2p}^{z_{2p}} & 0 \\ 
	0 & A^\dagger \gamma_{(p)}\Pi_{2p-1}^{z_{2p-1}}\Pi_{2p}^{z_{2p}}A
	\end{pmatrix}\,,
	\qquad (p=1,2,\cdots,d/2)\,,  \notag\\
&S_{(d+2)/2}^{z_{d+1}z_{d+2}}
  =	\begin{pmatrix}
	AA^\dagger \gamma^{d+1}\Pi_{d+1}^{z_{d+1}}\Pi_{d+1}^{z_{d+2}} & 0 \\ 
	 0 & -A^\dagger \gamma^{d+1}\Pi_{d+1}^{z_{d+1}}\Pi_{d+1}^{z_{d+2}}A
	\end{pmatrix}\,,
\end{align}
%
for $d=$ even and
%
\begin{align}
&S_p^{z_{2p-1}z_{2p}}
  =	\begin{pmatrix}
     AA^\dagger\gamma_{(p)}\Pi_{d(2p-1)}^{z_{2p-1}}\Pi_{d(2p)}^{z_{2p}} & 0 \\ 
     0 & A^\dagger  \gamma_{(p)}\Pi_{d(2p-1)}^{z_{2p-1}}\Pi_{d(2p)}^{z_{2p}} A
\end{pmatrix}\,,
	\qquad (p=1,2,\cdots,(d-1)/2)\,,  \notag\\
&S_{(d+1)/2}^{z_{d}z_{d+1}}
  =	\begin{pmatrix}
    iAA^\dagger \gamma^{y_d}\Pi_{d+1}^{z_{d}}\Pi_{d(d+1)}^{z_{d+1}} & 0 \\ 
    0 & -iA^\dagger \gamma^{y_d}\Pi_{d+1}^{z_{d}}\Pi_{d(d+1)}^{z_{d+1}}A
   \end{pmatrix}\,,
\end{align}
%
for $d=$ odd, where we introduce the p-th internal chirality $\gamma_{(p)}=i\gamma^{y_{2p-1}}\gamma^{y_{2p}}$\footnote{ The product of $\gamma_{(p)}\ (p=1,2,\cdots,\lfloor d\rfloor/2)$ equals to the internal chirality $\gamma^{d+1}$ for $d=$ even and $-i\gamma^{y_d} $ for $d=$ odd.}.
Since $S_p^{z_{2p-1}z_{2p}}$\ ($p=1,2,\cdots,\lfloor d/2\rfloor+1$)
commute with each other as well as $H$ and $Z_{i}$
($i=1,2,\cdots,2\lfloor d/2\rfloor+2$), eigenstates of $S_p^{z_{2p-1}z_{2p}}$
can become simultaneous eigenstates of $H$ and $Z_{i}$.
Furthermore, since $S_p^{z_{2p-1}z_{2p}}$ satisfy the relations
$\big(S_p^{z_{2p-1}z_{2p}}\big)^{2}=H^{2}$,
we can parameterize the eigenvalues of $S_p^{z_{2p-1}z_{2p}}$ as
%
\begin{align}
S_p^{z_{2p-1}z_{2p}}\Phi^{(n)}_{s_1\cdots s_p\cdots s_{\lfloor d/2\rfloor+1},\,\vec{z}\,}(y)
 =s_p\,m_{n}^{2}\,\Phi^{(n)}_{s_1\cdots s_p\cdots s_{\lfloor d/2\rfloor+1},\,\vec{z}\,}(y)
  \qquad (p=1,\cdots,\lfloor d/2\rfloor+1)\,,
\end{align}
%
with $s_{p}=\pm 1$ for the non-zero energy states \footnote{ For $d=$ even, $s_p\ (p=1,2,\cdots,d/2)$ correspond to the eigenvalues of the p-th internal chirality $\gamma_{(p)}$, and furthermore, $s_{(d+2)/2}$ corresponds to the eigenvalue of 4+d dimensional chirality since $S_{(d+2)/2}^{z_{d+1}z_{d+2}}$ are naively given by the product of $(-1)^F$ and $\gamma^{d+1}$. Thus the eigenvalues $s_p\ (p=1,2,\cdots,(d+2)/2)$ are independent.
The similar result is also obtained for $d=$ odd.
}.
Here, we have replaced the index $\alpha$ by $s_{1}s_{2}\cdots s_{\lfloor d/2\rfloor+1}$,
which denote the eigenvalues of $S_p^{z_{2p-1}z_{2p}}$\ 
($p=1,2,\cdots,\lfloor d/2\rfloor+1$).

From the relations
%
\begin{align}
&Q_{i}^{z_{i}} S_p^{z_{2p-1}z_{2p}}
 = \left\{\begin{matrix}
            -S_p^{z_{2p-1}z_{2p}}Q_{i}^{z_{i}}  & \text{for } i=2p-1, 2p\,, \\
            +S_p^{z_{2p-1}z_{2p}}Q_{i}^{z_{i}} & \text{for } i\ne2p-1, 2p\,, \\
          \end{matrix}
          \right. \\
&[\,Q_{i}^{z_{i}}\,,\,H\,]
  = [\,Q_{i}^{z_{i}}\,,\,Z_{j}\,]
  = 0\,, \qquad (i,j=1,2,\cdots,2\lfloor d/2\rfloor+2)\,,
\end{align}
%
we find that the supercharges $Q_{2p-1}^{z_{2p-1}}$
(or $Q_{2p}^{z_{2p}}$)\ ($p=1,2,\cdots,\lfloor d/2\rfloor+1$)\footnote{
Since we are considering the eigenstates of 
$S_p^{z_{2p-1}z_{2p}}= -iQ_{2p-1}^{z_{2p-1}} Q_{2p}^{z_{2p}}$,
the action of $Q_{2p}^{z_{2p}}$ on 
$\Phi^{(n)}_{s_1\cdots s_p\cdots s_{\lfloor d/2\rfloor+1},\,\vec{z}\,}(y)$
is essentially equivalent to that of $Q_{2p-1}^{Z_{2p-1}}$.
}
flips the sign of the eigenvalues of $S_p^{z_{2p-1}z_{2p}}$ but do not
change other eigenvalues.
This fact implies that the set of 
$\{ \Phi^{(n)}_{s_1\cdots s_p\cdots s_{\lfloor d/2\rfloor+1},\,\vec{z}\,}(y)\ 
\text{with}\ s_{p}=\pm1\ (p=1,2,\cdots,\lfloor d/2\rfloor+1) \}$
is $2^{\lfloor d/2\rfloor+1}$-fold degenerate and form a supermultiplet
of the $\mathcal{N}=4\lfloor d/2\rfloor+4$ extended QM SUSY algebra
with the central charges.
Actually, we can explicitly construct the supermultiplet from 
$\Phi^{(n)}_{++\cdots+,\,\vec{z}\,}(y)$ as
%
\begin{align}
\Phi^{(n)}_{s_{1}s_{2}\cdots s_{\lfloor d/2\rfloor+1},\,\vec{z}\,}(y)
 =\frac{1}{(m_n)^{s}}
   \left(Q_1^{z_1}\right)^{(1-s_{1})/2}
   \left(Q_3^{z_3}\right)^{(1-s_{2})/2}\cdots
   \left(Q_{2\lfloor d/2\rfloor+1}^{z_{2\lfloor d/2\rfloor+1}}\right)^{
   (1-s_{\lfloor d/2\rfloor+1})/2}\Phi^{(n)}_{++\cdots +,\,\vec{z}\,}(y)\,,
   \label{eq:1/2BPS_state}
\end{align}
%
where $s=\frac{1}{2}(1-s_{1})+ \cdots + \frac{1}{2}(1-s_{2\lfloor d/2\rfloor+1})$.

As we have seen so far, the number of the non-trivial supercharges
reduces to half, and the $2^{\lfloor d/2\rfloor+1}$-fold degenerate
states for fixed $m_{n}$ and $\vec{z}$ are related by the reduced 
$2\lfloor d/2\rfloor+2$ supercharges.
This situation is known as a short representation in the context of 
extended supersymmetry with central charges, and the eigenstates 
\eqref{eq:1/2BPS_state} are called the 1/2-BPS states 
\cite{Witten:1978mh,WessBagger}.

\section{Examples}
\label{sec:examples}

In this section, we examine the models with the hyperrectangle and the torus
extra dimensions, which realize the $\mathcal{N}=4\lfloor d/2\rfloor+4$ 
extended QM SUSY, and confirm the results given in the previous section.

\subsection{Hyperrectangle}
\label{subsec:hyperrectangle}

Let us consider the example of the action \eqref{eq:action} whose extra dimensional space $\Omega$ is given by the $d$-dimensional hyperrectangle,
%
\begin{align}
	 \Omega= \bigg[-\frac{L_1}{2},\frac{L_1}{2}\bigg]\times\cdots\times\bigg[-\frac{L_d}{2},\frac{L_d}{2}\bigg]\,,
\end{align}
%
where $L_{k}$\ ($k=1,2,\cdots,d$) is the length of the $k$-th side of the
hyperrectangle with the Dirichlet boundary condition imposed on the left-handed KK mode functions,
%
\begin{align}
	 \fbd_{s_{1}\cdots s_{\lfloor d/2\rfloor}}^{(n)}(y) = 0
	 \quad \text{at}\ y_{k}=-\frac{L_k}{2}, \frac{L_k}{2}\ \ \ (k=1,\cdots,d)\ .
	 \label{eq:BC_rec}
\end{align}
%
This boundary condition satisfies the requirement \eqref{eq:non-local_BC}, 
and we can confirm that all the supercharges $Q^{\pm}_{i}$ are Hermitian 
and well-defined.

The $n$-th KK mode functions $\fbd_{s_{1}\cdots s_{\lfloor d/2\rfloor}}^{(n)}(y)\ ,\ \gbd_{s_{1}\cdots s_{\lfloor d/2\rfloor}}^{(n)}(y)$ with KK mass $m_n^2 >0$ are 
found to be written into the form
%
\begin{align}
	 &\fbd_{s_{1}\cdots s_{\lfloor d/2\rfloor}}^{(n)}(y)=h^{(n)}(y)\,
	  \ebd_{s_{1}\cdots s_{\lfloor d/2\rfloor}}\,,
	  &&\gbd_{s_{1}\cdots s_{\lfloor d/2\rfloor}}^{(n)}(y)=  \frac{1}{m_{n}} A^\dagger h^{(n)}(y)\,
	   \ebd_{s_{1}\cdots s_{\lfloor d/2\rfloor}}\,,
\end{align}
%
where the second relation comes from the first equation in \eqref{eq:SUSY_relation}.
The scalar function $h^{(n)}(y)$ and the mass eigenvalue $m_{n}$ are given by
%
\begin{align}
	 h^{(n)}(y) & 
	 =\prod_{k=1}^{d}
	 \sqrt{\frac{2}{L_{k}}}\sin\left(\frac{n_{k}\pi}{L_{k}}
	 \Big(y_{k}+\frac{L_{k}}{2}\Big)\right)\,,\\
	 m_{n}^{2} & =M^{2}+\sum_{k=1}^{d}\bigg(\frac{n_{k}\pi}{L_{k}}\bigg)^{2}\,, \ \ \ \ 
	 (n_{k}=1,2,\cdots;\ \ k=1,\cdots,d)\,,
\end{align}
%
The $\ebd_{s_{1}\cdots s_{\lfloor d/2\rfloor}}$ indicate the basis vectors of the spinor space, and are chosen as eigenvectors of $\gamma_{(p)}=i\gamma^{y_{2p-1}}\gamma^{y_{2p}}$ \cite{Polchinski1-String,Polchinski2-String}:
%
\begin{align}
	\gamma_{(p)}\boldsymbol{e}_{s_{1} \cdots s_{p} \cdots s_{{\lfloor d/2\rfloor}}}
	& = s_{p}\boldsymbol{e}_{s_{1} \cdots s_{p} \cdots s_{{\lfloor d/2\rfloor}}}\,,
\end{align}
%
for all $p=1,2,\cdots,\lfloor d/2\rfloor$, where $s_p=\pm1$ represents an eigenvalue of the $p$-th internal chirality {of $\gamma_{(p)}$}.
To fix the normalization factors of $\ebd_{s_{1}\cdots s_{\lfloor d/2\rfloor}}$, 
we define the $\boldsymbol{e}_{s_{1}\cdots s_{{\lfloor d/2\rfloor}}}$ from $\boldsymbol{e}_{+\cdots +}$ as
(c.f. Eqs.~(\ref{eq:Qpm_even}), (\ref{eq:Qpm_odd}) and (\ref{eq:1/2BPS_state}))
%
\begin{align}
	\boldsymbol{e}_{s_{1}s_{2}\cdots s_{{\lfloor d/2\rfloor}}}=
	\begin{cases}
	\left(i\gamma^{d+1}\gamma^{y_1}\right)^{(1-s_{1})/2}
	\left(i\gamma^{d+1}\gamma^{y_3}\right)^{(1-s_{2})/2}\cdots
	\left(i\gamma^{d+1}\gamma^{y_{ d-1}}\right)^{(1-s_{d/2})/2}
	\boldsymbol{e}_{++ \cdots +} & (d=\text{even})\,, \\
	\left(-\gamma^{y_d}\gamma^{y_1}\right)^{(1-s_{1})/2}
	\left(-\gamma^{y_d}\gamma^{y_3}\right)^{(1-s_{2})/2}\cdots
	\left(-\gamma^{y_d}\gamma^{y_{ d-2}}\right)^{(1-s_{(d-1)/2})/2}
	\boldsymbol{e}_{++ \cdots +} & (d=\text{odd}) \,.
	\end{cases}
\end{align}
%

Then, we can construct the eigenfunctions of the model as follows:
%
\begin{align}
\Phi_{s_{1}\cdots s_{{\lfloor d/2\rfloor}}s_{{\lfloor d/2\rfloor+1}},\,\vec{z}\,}^{(n)}(y)
 = \left\{
     \begin{matrix}
        \begin{pmatrix}
	      \boldsymbol{f}_{s_{1}\cdots s_{{\lfloor d/2\rfloor}}}^{(n)}(y) \\ 0
	    \end{pmatrix} 
	     \qquad \text{for}\ s_{{\lfloor d/2\rfloor+1}}=s_{1}s_{2}\cdots s_{\lfloor d/2\rfloor}\,,
            \\
        \hspace{2mm}
        \begin{pmatrix}
	      0 \\ \boldsymbol{g}_{s_{1}\cdots s_{{\lfloor d/2\rfloor}}}^{(n)}(y)
	    \end{pmatrix} 
	     \qquad \text{for}\ s_{{\lfloor d/2\rfloor+1}}=-s_{1}s_{2}\cdots s_{\lfloor d/2\rfloor}\,,
     \end{matrix}
   \right.
	\label{eq:rec_eigenfunction}
\end{align}
%
where 
$\vec{z}=(z_{1},z_{2},\cdots,z_{d})$ is given by
%
\begin{align}
	\vec{z}=\begin{cases}
	\Big((-)^{n_1+1},\cdots,(-)^{n_d+1}\Big) & (d=\text{even})\,,\\
	\Big((-)^{n_1+n_d},\cdots,(-)^{n_{d-1}+n_d},(-)^{n_d+1}\Big) & (d=\text{odd})\,.
	\end{cases}
	\label{eq:z_01}
\end{align}
%

Then, we can show that the eigenfunctions \eqref{eq:rec_eigenfunction}
satisfy the same relations as \eqref{eq:1/2BPS_state} and form the 
supermultiplet of the $\mathcal{N}=4\lfloor d/2\rfloor+4$ extended QM SUSY,
as the 1/2-BPS states.
Since the eigenvalues of $Z_{i}$ are unique at each KK level as shown in 
\eqref{eq:z_01}, the degeneracy of the 4D spectrum at each KK level is 
$2^{\lfloor d/2\rfloor+1}$ and the eigenfunctions \eqref{eq:rec_eigenfunction}
are mutually related by the supercharges $Q_{i}^{z_{i}}$\ 
($i=1,2,\cdots,2\lfloor d/2\rfloor+2$) at each KK level.
It is interesting to point out that the KK mode functions 
$\boldsymbol{g}_{s_{1}\cdots s_{{\lfloor d/2\rfloor}}}^{(n)}(y)$
are not eigenfunctions of the reflection operators $R_{k}\ (k=1,2,\cdots,d)$
and $P$, although $\boldsymbol{f}_{s_{1}\cdots s_{{\lfloor d/2\rfloor}}}^{(n)}(y)$
are eigenfunctions of them.
On the other hand, 
$\Phi_{s_{1}\cdots s_{{\lfloor d/2\rfloor}}s_{{\lfloor d/2\rfloor+1}},\,\vec{z}\,}^{(n)}(y)$
are eigenfunctions of the central charges $Z_{i}\ (i=1,2,\cdots,2\lfloor d/2\rfloor+2)$.
Thus, $Z_{i}$ can be regarded as \lq\lq reflection\rq\rq\ operators compatible with
the supersymmetry, as noticed in the previous section.
%
%

\subsection{Torus}
\label{subsec:torus}

Next, we consider the model that the extra dimensional space $\Omega$ is 
given by the $d$-dimensional torus,
%
\begin{align}
	\Omega= \bigg[-\frac{L_1}{2},\frac{L_1}{2}\bigg)\times\cdots\times\bigg[-\frac{L_d}{2},\frac{L_d}{2}\bigg)
\end{align}
%
with the periodic boundary condition for KK mode functions,
%
\begin{align}
\begin{aligned}
	&\fbd_{s_{1}\cdots s_{\lfloor d/2\rfloor},\,\vec{z}^{\,\prime}}^{(n)}(y_1,\cdots,y_k+L_k,\cdots y_d)=\fbd_{s_{1}\cdots s_{\lfloor d/2\rfloor},\,\vec{z}^{\,\prime}}^{(n)}(y_1,\cdots,y_k,\cdots, y_d)\,,\\
	&\gbd_{s_{1}\cdots s_{\lfloor d/2\rfloor},\,\vec{z}^{\,\prime}}^{(n)}(y_1,\cdots,y_k+L_k,\cdots y_d)=\gbd_{s_{1}\cdots s_{\lfloor d/2\rfloor},\,\vec{z}^{\,\prime}}^{(n)}(y_1,\cdots,y_k,\cdots, y_d)\,,
	\ \ \ (k=1,\cdots,d)\ .
	\label{eq:BC_torus}
\end{aligned}
\end{align}
%
The above periodic boundary condition satisfies the requirement \eqref{eq:non-local_BC}, 
and all the supercharges are shown to be Hermitian and well-defined.

Then, the $n$-th KK mode functions $\fbd_{s_{1}\cdots s_{\lfloor d/2\rfloor},\,\vec{z}^{\,\prime}}^{(n)}(y)\ ,\ \gbd_{s_{1}\cdots s_{\lfloor d/2\rfloor},\,\vec{z}^{\,\prime}}^{(n)}(y)$ with KK mass $m_n$
are found to be of the form
%
\begin{align}
	&\fbd_{s_{1}\cdots s_{\lfloor d/2\rfloor},\,\vec{z}^{\,\prime}}^{(n)}(y)
	=h^{(n)}_{\vec{z}^{\,\prime}}(y)\,\ebd_{s_{1}\cdots s_{\lfloor d/2\rfloor}}\,,
	\ \ \ \ \gbd_{s_{1}\cdots s_{\lfloor d/2\rfloor},\,\vec{z}^{\,\prime}}^{(n)}(y)
	=  \frac{1}{m_{n}} A^\dagger h^{(n)}_{\vec{z}^{\,\prime}}(y)\,
	\ebd_{s_{1}\cdots s_{\lfloor d/2\rfloor}}\,,
	\\
	&h^{(n)}_{\vec{z}^{\,\prime}}(y) 
	=\prod_{k=1}^{d}h^{(n_k)}_{z_k'}(y)\,,\ \ \ \  
	h^{(n_k)}_{z_k'}(y)=
	\begin{cases}
	\sqrt{\frac{2}{L_{k}}}\cos\left(\frac{2n_{k}\pi}{L_{k}}
	y_{k}\right)&(z_k'=+1)\,,
	\\
	\sqrt{\frac{2}{L_{k}}}\sin\left(\frac{2n_{k}\pi}{L_{k}}
	y_{k}\right)&(z_k'=-1)\,,
	\end{cases}
	\\
	&m_{n}^{2}=M^{2}+\sum_{k=1}^{d}\bigg(\frac{2n_{k}\pi}{L_{k}}\bigg)^{2}\,,
	\ \ \ \ (n_{k}=0,1,2,\cdots;\ \ k=1,\cdots, d)\,,
\end{align}
%
where $z_k'=\pm1$ and $\vec{z}^{\,\prime} = (z_1',\cdots,z_d')$.

Then, we can construct the eigenfunctions of the model as 
%
\begin{align}
\Phi_{s_{1}\cdots s_{{\lfloor d/2\rfloor}}s_{{\lfloor d/2\rfloor+1}},\,\vec{z}\,}^{(n)}(y)
 = \left\{
     \begin{matrix}
        \begin{pmatrix}
	      \boldsymbol{f}_{s_{1}\cdots s_{{\lfloor d/2\rfloor}},\,\vec{z}^{\,\prime}}^{(n)}(y) \\ 0
	    \end{pmatrix} 
	     \qquad \text{for}\ s_{{\lfloor d/2\rfloor+1}}=s_{1}s_{2}\cdots s_{\lfloor d/2\rfloor}\,,
            \\
        \hspace{2mm}
        \begin{pmatrix}
	      0 \\ \boldsymbol{g}_{s_{1}\cdots s_{{\lfloor d/2\rfloor}},\,\vec{z}^{\,\prime}}^{(n)}(y)
	    \end{pmatrix} 
	     \qquad \text{for}\ s_{{\lfloor d/2\rfloor+1}}=-s_{1}s_{2}\cdots s_{\lfloor d/2\rfloor}\,,
     \end{matrix}
   \right.
	\label{eq:torus_eigenfunction}
\end{align}
%
where 
$\vec{z}$ is given by
%
\begin{align}
	\vec{z}=\begin{cases}
	(z_1',\cdots,z_d') & (d=\text{even})\,,\\
	(z_d'z_1',\cdots,z_d'z_{d-1}',z_d') & (d=\text{odd})\,.
	\end{cases}
\end{align}
%

Then, we can show that the eigenfunctions \eqref{eq:torus_eigenfunction} satisfy
the same relation as \eqref{eq:1/2BPS_state} and form the supermultiplet of the
$\mathcal{N} = 4\lfloor d/2\rfloor +4$ extended QM SUSY, as the $1/2$-BPS states.
However, unlike the hyperrectangle case, both of the eigenstates with
$z_{k}=+1$ and $-1$ for $n_{k}\ne 0$ ($k=1,2,\cdots,d$) are degenerate in the
4D mass spectrum.
This implies that the additional degeneracy $2^{d-N_{0}}$ appears in the 4D
spectrum, where $N_{0}$ is the number of zeros in 
$\{n_{1}, n_{2}, \cdots, n_{d} \}$\footnote{
Note that when $n_{k}=0$, $h^{(n_{k})}_{z_{k}'}(y)$ for $z_{k}'=-1$ is trivial, i.e.
$h^{(0)}_{z_{k}'}(y) = 0$.
Thus, there is no degeneracy in $h^{(n_{k})}_{z_{k}'}(y)$ for $n_{k}=0$.
}.
The origin of the degeneracy comes from the extra degrees of
freedom with respect to the parity even or odd for each reflection: $y_{k} \to -y_{k}$ 
in $h^{(n)}_{\vec{z}^{\,\prime}}(y)$.
Therefore, the 4D mass spectrum is 
$(2^{2\lfloor d/2\rfloor+1} \times 2^{d-N_{0}})$-fold degenerate for 
the KK modes labeled by $\{n_{1}, n_{2}, \cdots, n_{d} \}$.
%
%

\section{Summary and discussion}
\label{sec:summary}

In this paper, we have revealed that the $\mathcal{N}$-extended QM SUSY with 
the central charges is hidden in the 4D mass spectrum of the higher dimensional Dirac action.
The supercharges are obtained as the extension of the $\Ncal=2$ QM SUSY based on the algebraic properties of the internal gamma matrices and the reflection symmetries of the extra dimensions.
The central charges are interpreted as the supersymmetric extension of the reflection operators.

We have also examined the representation of the extended supersymmetry algebra 
and found that the supermultiplet corresponds to the short multiplet of the 
1/2-BPS states.
Furthermore, we have explicitly confirmed that the KK mode functions in the models
of the hyperrectangle and the torus extra dimensions can be properly classified
by the representations of the $\mathcal{N}=4\lfloor d/2 \rfloor+4$ extended
QM SUSY algebra with the central charges.

In this paper, we have restricted to the cases that all the $4\lfloor d/2 \rfloor+4$ 
supercharges are well-defined. 
Other boundary conditions, other extra dimensions and non-trivial background fields 
would break (or partially break) the extended supersymmetry.
For example, if there are no reflection symmetries in extra dimensions, 
the geometric supercharges \eqref{eq:geo_even}, \eqref{eq:geo_odd} 
(and also the supercharges \eqref{eq:redef_supercharge})
become ill-defined, although the algebraic supercharges can be well-defined 
in this case with suitable boundary conditions. 
Therefore, it would be of great importance to clarify how the extended supersymmetry
found in this paper is broken by the choice of boundary conditions, 
extra dimensional spaces and background fields.

It is interesting to note that there are possibilities that further structures 
might be hidden in the 4D mass spectrum in general settings. 
The central charges in our models result from the symmetries of the extra dimensions.
Thus, we can expect that new types of central charges will 
appear in models with other symmetries.
Furthermore, since it is known that central charges are closely related to topological properties \cite{Witten:1978mh,Seiberg:1994rs,Seiberg:1994aj},
it is also interesting to investigate models of curved extra dimensions
or background fields with non-trivial topologies, e.g. sphere, soliton, magnetic flux, etc.

In addition, since we have obtained the new extended supersymmetry with the central charges, 
it would be worthwhile to search for new types of exactly solvable models 
by use of this supersymmetry.
The issues mentioned above remain to be done in future works.

\section*{Acknowledgement}

This work is supported
in part by Grants-in-Aid for Scientific Research [No.~18K03649~(Y.F. and M.S.)]
from the Ministry of Education, Culture, Sports, Science and Technology (MEXT) in Japan.
K.N. had been funded by the research grant (No.~PG052102)
from Korea Institute for Advanced Study~(KIAS),
and has been supported by
the European Union through the European Regional Development Fund -- the Competitiveness and 
Cohesion Operational Programme (KK.01.1.1.06),
the European Union's Horizon 2020 research and
innovation program under the Twinning grant agreement No.~692194, RBI-T-WINNING,
and the grant funded from the European Structural and Investment Funds,
RBI-TWINN-SIN.

\bibliographystyle{elsarticle-num}
\bibliography{Reference}

\end{document}